\documentclass[twocolumn,aps]{revtex4}
\begin{document}
\draft
\title{On singular Lagrangian underlying the Schr\"odinger equation.}
\author{A.A. Deriglazov$^{(*)}$}
\address{Dept. de Matematica, ICE, Universidade Federal de Juiz de Fora,
MG, Brasil.}
\date{}

\begin{abstract}
We analyze the properties that manifest Hamiltonian nature of the Schr\"odinger equation and show that it can be considered as originating from singular Lagrangian action (with two second class constraints presented in the Hamiltonian formulation). It is used to show that any solution to the Schr\"odinger equation with time independent potential can be presented in the form $\Psi=(-\frac{\hbar^2}{2m}\triangle+V)\phi+i\hbar\partial_t\phi$, where the real field $\phi(t, x^i)$ is some solution to nonsingular Lagrangian theory being specified below. Preservation of probability turns out to be the energy conservation law for the field $\phi$. After introducing the field into the formalism, its mathematical structure becomes analogous to those of electrodynamics: the real field $\phi$ turns out to be a kind of potential for a wave function.
\end{abstract}

\maketitle
\noindent
%{\bf DOI:}
%{\bf PACS numbers:} 11.10.Ef, 03.65.Ca
%{\bf Keywords:} Constrained theories, Schr\"odinger equation
\section{Introduction.}
Hamiltonian character of the Schr\"odinger equation is widely explored in various quantum mechanical applications [1-6].
In classical mechanics, Hamiltonian equations for the phase space variables $q, p$ normally  originate from a Lagrangian formulation for the configuration variables $q$: there exists an action that implies the second order equations equivalent to the Hamiltonian ones. It is the aim of this work to show that the Schr\"odinger equation with time independent potential admits a similar treatment.

In fact, the problem has been raised already by Schr\"odinger [7]. Eq. (\ref{-48S7}) below has been tested by Schr\"odinger as a candidate for the wave function equation and then abandoned. So, the real field $\phi$ appeared in this equation will be called the Schr\"odinger field. In Section 2 we establish the simple formula (\ref{-48S6}) that generates solutions to the Schr\"odoinger equation from solutions to the Schr\"odinger field equation. In Section 3 we present singular Lagrangian theory that implies unified description for both the Schr\"odinger equation and the Schr\"odinger field equation. The unified formulation is used, in particular, to prove that any solution to the Schr\"odinger equation can be presented according to the formula (\ref{-48S6}).
It implies, that after introduction the Schr\"odinger field into the formalism, its mathematical structure becomes analogous to those of electrodynamics. In particular, as well as $A^\mu$ represents a potential for magnetic and electric fields, the Schr\"odinger field turns out to be the wave function potential, giving its real and imaginary parts according to Eq. (\ref{-48S6}). Other similarities are summarized in the Conclusion.

\section{Nonsingular Lagrangian associated with the Schr\"odinger equation}
We restrict ourselves to the one-particle Schr\"odinger equation with time independent potential $V(x^i)$
\begin{eqnarray}\label{-48S1}
i\hbar\dot\Psi=-\triangle\Psi+V\Psi.
\end{eqnarray}
We use the notation $\triangle$$=$$\frac{\hbar^2}{2m}\frac{\partial^2}{\partial x^{i2}}$,
$\vec\nabla$$=$$\frac{\hbar}{\sqrt{2m}}\frac{\partial}{\partial x^i}$,  $\dot\varphi$$=$$\frac{\partial\varphi(t, x^i)}{\partial t}$.
It is equivalent to the system of two equations for two real functions (we take real and imaginary parts of the wave function $\Psi(t, x^i)$, $\Psi$$=$$\varphi$$+$$ip$). One obtains
\begin{eqnarray}\label{-48S2}
\hbar\dot\varphi=
-\left(\triangle-V\right)p,
\end{eqnarray}
\begin{eqnarray}\label{-48S22}
\hbar\dot p=\left(\triangle-V\right)\varphi. \quad
\end{eqnarray}
Considering $p(t, x^i)$ as conjugate momenta for the field $\varphi(t, x^i)$, the system  has the Hamiltonian form,  $\dot\varphi$$=$$\{\varphi, H\}$, $\dot p $$=$$\{p, H\}$, where $\{ , \}$ stands for the Poisson bracket and $H$ is the Hamiltonian
\begin{eqnarray}\label{-48S3}
H=\frac{1}{2\hbar}\int d^3x[\vec\nabla\varphi\vec\nabla\varphi+
\vec\nabla p\vec\nabla p+
V(\varphi^2+p^2)].
\end{eqnarray}
Hence the equations (\ref{-48S2}), (\ref{-48S22}) arise from variation problem for Hamiltonian action obtained according to the known rule
\begin{eqnarray}\label{-48S4}
S_H=\int dt d^3x\left[p\dot\varphi-H\right]= \qquad \qquad \qquad \cr
\int dt d^3x\left[\frac{i\hbar}{2}(\Psi^*\dot\Psi-
\dot\Psi^*\Psi)-\vec\nabla\Psi^*\vec\nabla\Psi-V\Psi^*\Psi\right].
\end{eqnarray}
Following the classical mechanics prescription, to construct the corresponding Lagrangian formulation (if any) one needs to resolve Eq. (\ref{-48S2}) with respect to $p$ and then to substitute the result either in Eq. (\ref{-48S22}) or into the Hamiltonian action (\ref{-48S4}).
It leads immediately to rather formal nonlocal expression $p$$=$$-\hbar(\triangle-V)^{-1}$$\dot\varphi$. So, the Schr\"odinger system can not be obtained starting from some (nonsingular) Lagrangian. Nevertheless, there exists nonsingular Lagrangian field theory with the property that any solution to the Schr\"odinger equation can be constructed from some solution to this theory. To find it let us look for solutions of the form
\begin{eqnarray}\label{-48S6}
\Psi=-(\triangle-V)\phi+i\hbar\dot\phi,
\end{eqnarray}
where $\phi(t, x^i)$ is some real function. $\Psi$ will be solution to the Schr\"odinger equation if $\phi$ obeys the equation
\begin{eqnarray}\label{-48S7}
\hbar^2\ddot\phi+(\triangle-V)^2\phi=0,
\end{eqnarray}
the latter follows from the Lagrangian action
\begin{eqnarray}\label{-48S8}
S[\phi]=\int dtd^3x\left[\frac{\hbar}{2}\dot\phi\dot\phi-
\frac{1}{2\hbar}[(\triangle-V)\phi]^2\right].
\end{eqnarray}
It is considered here as the classical theory of field $\phi$ on the given external background $V(x^i)$. The action involves the Planck's constant as a parameter. After the rescaling $(t, x^i, \phi)$$\rightarrow$$(\hbar t, \hbar x^i, \sqrt\hbar\phi)$ it appears in the potential only, $V(\hbar x ^i)$, and thus plays the role of coupling constant of the field $\phi$ with the background.

According to the formula (\ref{-48S6}), both probability density and phase of a wave function can be presented through the Schr\"odinger field. Taking $\Psi$$=$$\sqrt P\exp{\frac{i}{\hbar}S}$ one obtains \begin{eqnarray}\label{-48S8.1}
P=\hbar^2(\dot\phi)^2+[(-\triangle+V)\phi]^2=2\hbar E, \cr
S=-\hbar\arctan\sqrt{\frac{T}{U}}, \qquad \qquad \qquad \quad
\end{eqnarray}
where $E$$=$$T+U$ is energy density of the Schr\"odinger field. The first equation states that probability density is the energy density of the Schr\"odinger field. Invariance of the action under the time translations implies the current equation
\begin{eqnarray}\label{-48S8.1}
\partial_tE+\vec\nabla(2\hbar^{-2}E\vec\nabla S)=0.
\end{eqnarray}
Thus preservation of probability is just the energy conservation law of the theory (\ref{-48S8}).

It is instructive to compare Hamiltonian equations of the theory (\ref{-48S8})
\begin{eqnarray}\label{-48S9}
\hbar\dot\phi=p, \qquad
\hbar\dot p=-(\triangle-V)^2\phi,
\end{eqnarray}
with the Schr\"odinger system. One notices the following correspondence  among solutions to these systems: a) If the functions $\varphi, p$ obey Eq. (\ref{-48S2}), (\ref{-48S22}), then the functions $\phi\equiv\varphi$,
$-(\triangle-V)p$ obey Eq. (\ref{-48S9}). b) If the functions $\phi, p$ obey Eqs. (\ref{-48S9}), then $\varphi\equiv-(\triangle-V)\phi$, $p$ obey
Eqs. (\ref{-48S2}), (\ref{-48S22}). Kernel of the map $(\varphi, p)$$\rightarrow$$(\phi, p)$ is composed by pure imaginary time independent wave functions $\Psi$$=$$i\Pi(x^i)$, where $\Pi$ is any solution to the stationary Schr\"odinger equation $(\triangle-V)\Pi=0$.

Any solution to the field theory (\ref{-48S8}) determines some solution to the Schr\"odinger equation according to Eq. (\ref{-48S6}). Then one should ask whether an arbitrary solution to the Schr\"odinger equation can be presented in the form (\ref{-48S6})?
An affirmative answer will be obtained~\footnote{Passage from classical to quantum mechanics is achieved by a quantization procedure. Eq. (\ref{-48S6}) implies the passage $\Psi$$\rightarrow$$\phi$ from quantum mechanics of a particle to some classical field theory,
so the procedure can be called "dequantization". Combining the two procedures, with the classical mechanics is associated the classical field. Solving the classical field theory one is able to describe quantum behavior of the particle.} in the next section using the Dirac approach to description  the constrained systems [8-11]. Besides, in this setting one obtains
more systematic treatment of the observations made above: there exists the singular Lagrangian theory subject to second class constraints underlying both the Schr\"odinger equation and the classical theory (\ref{-48S8}).

To motivate our appeal to the constrained theories, let us return to the Schr\"odinger system (\ref{-48S2}), (\ref{-48S22}). Its treatment as a Hamiltonian system does not allow one to construct the corresponding Lagrangian formulation owing to presence the spatial derivatives of momenta in the Hamiltonian. To avoid the problem, let us try to treat the Schr\"odinger system as a generalized Hamiltonian system. Namely, one rewrites (\ref{-48S2}), (\ref{-48S22}) in the form
\begin{eqnarray}\label{-48S10}
\dot\varphi=\{\varphi, H'\}', \qquad
\dot p=\{p, H'\}',
\end{eqnarray}
where $H'$ is the "free field" generalized Hamiltonian\footnote{In classical mechanics, inclusion of an interaction into a symplectic structure of the phase manifold has been investigated by Souriau [12]. It is discussed also in the framework of non commutative theories [13].}
\begin{eqnarray}\label{-48S11}
H'=\int d^3x\frac{1}{2\hbar}(p^2+\varphi^2)=\int d^3x\frac{1}{2\hbar}\Psi^*\Psi,
\end{eqnarray}
and the non canonical Poisson bracket is specified by\footnote{Similar construction is known for the Maxwell equations. They have been recognized as the generalized Hamiltonian equations in [14, 15].}
\begin{eqnarray}\label{-48S12}
\{\varphi, \varphi\}'=\{p, p\}'=0, \qquad \qquad \qquad \cr
\{\varphi(t, x), p(t, y)\}'=-(\triangle-V)\delta^3(x-y).
\end{eqnarray}
In contrast to $H$, the Hamiltonian $H'$ does not involve the spatial derivatives of momentum.

Non canonical bracket turns out to be a characteristic property of the theories with second class constraints. In this case
the constraints can be taken into account by transition from the Poisson to the Dirac bracket, the latter represents an example of non canonical bracket~\footnote{It should be noticed that being restricted to the dynamical sector variables, the Dirac bracket is non degenerated.}. Hamiltonian equations for dynamical variables, being written in terms of the Dirac bracket, form a  generalized Hamiltonian system. So, equations (\ref{-48S10})-(\ref{-48S12}) represent a hint to search for  associated constrained Lagrangian.

\section{Singular Lagrangian underlying the Schr\"odinger equation}
Here we obtain (\ref{-48S2}), (\ref{-48S22}) as Hamiltonian equations corresponding the Lagrangian theory
\begin{eqnarray}\label{-48S13}
S[\phi, \varphi]=\int dtd^3x\left[\frac{\hbar}{2}\dot\phi\dot\phi+\frac{1}{2\hbar}\varphi^2+
\frac{1}{\hbar}\varphi(\triangle-V)\phi\right],
\end{eqnarray}
written for two real fields $\phi(t, x^i)$, $\varphi(t, x^i)$ on a given external background $V(x^i)$. It implies the Lagrangian equations
\begin{eqnarray}\label{-48S13.1}
\hbar^2\ddot\phi-(\triangle-V)\varphi=0, \qquad
\varphi=-(\triangle-V)\phi.
\end{eqnarray}
As a consequence, both $\phi$ and $\varphi$ obey the second order equation (\ref{-48S7}). After the shift $\tilde\varphi\equiv\varphi+(\triangle-V)\phi$, the action acquires the form $S[\phi, \varphi]$$=$$S[\phi]$$+$$\frac{1}{2\hbar}\int\tilde\varphi^2$. Hence in this parametrization the fields $\phi$ and $\tilde\varphi$ decouple, and the only dynamical variable is $\phi$. Its evolution is governed by Eq. (\ref{-48S7}). Being rather natural, it is not unique possible parametrization of dynamical sector. To find another relevant parametrization, we would like to construct Hamiltonian formulation of the theory. One introduces the conjugate momenta $p$, $\pi$ for the fields $\phi$, $\varphi$ and defines their evolution according to
\begin{eqnarray}\label{-48S14}
p=\frac{\partial L}{\partial\dot\phi}=\hbar\dot\phi, \qquad
\pi=\frac{\partial L}{\partial\dot\varphi}=0.
\end{eqnarray}
The second equation does not contain time derivative of the fields, hence it represents  primary constraint of the theory. Then the Hamiltonian is
\begin{eqnarray}\label{-48S15}
H=\int d^3x\left[\frac{1}{2\hbar}(p^2-\varphi^2)-
\frac{1}{\hbar}\varphi(\triangle-V)\phi+v\pi\right],
\end{eqnarray}
where $v$ stands for the Lagrangian multiplier of the constraint. Preservation in time of the primary constraint, $\dot\pi$$=$$\{\pi, H\}$$=$$0$, implies the secondary one
$\varphi$$+$$(\triangle-V)\phi$$=$$0$. In turn, its preservation in time determines the Lagrangian multiplier
$v$$=$$-\frac{1}{\hbar}(\triangle-V)p$.
Hence the Dirac procedure stops on this stage. Evolution of the phase space variables is governed by the Hamiltonian equations
\begin{eqnarray}\label{-48S18}
\dot\phi=\frac{1}{\hbar}p, \qquad \dot p=\frac{1}{\hbar}(\triangle-V)\varphi, \cr
\dot\varphi=v\approx-\frac{1}{\hbar}(\triangle-V)p, \qquad \dot\pi\approx 0,
\end{eqnarray}
and by the constraints
\begin{eqnarray}\label{-48S19}
\pi=0, \qquad
\varphi+(\triangle-V)\phi=0.
\end{eqnarray}
The system implies that both $\phi$ and $\varphi$ obey Eq. (\ref{-48S7}).
Computing Poisson bracket of the constraints, one obtains on-shell non vanishing result,
$\{\varphi$$+$$(\triangle-V)\phi$$, \pi\}$$=$$\delta^3(x-y)$. According to the Dirac terminology, the constraints form a second class system.

We reminded the Dirac prescription for dealing with second class constraints. They are used to determine a part of variables in terms of others. The variables that have been thus determined are conventionally called non dynamical variables. Evolution of the remaining dynamical variables is governed by equations of first order with respect to time. They are obtained from the initial equations (\ref{-48S18}) taking into account the constraints as well as equations for the Lagrangian multipliers.

There is an equivalent way to obtain equations of motion for dynamical variables. One can write the Hamiltonian (\ref{-48S15}) in terms of dynamical variables, and to construct the Dirac bracket corresponding to the constraints (\ref{-48S19})
\begin{eqnarray}\label{-48S19.1}
\{A, B\}_D=\{A, B\}-\{A, \pi\}\{\varphi+(\triangle-V)\phi, B\} \cr
+\{A, \varphi+(\triangle-V)\phi\}\{\pi, B\}.
\end{eqnarray}
It implies $\{\pi, A\}_D$$=$$0$, $\{\phi, \varphi\}_D$$=$$0$, as well as
\begin{eqnarray}\label{-48S20}
\{\phi, p\}_D=\delta^3(x-y), \qquad
\{\phi, \phi\}_D=\{p, p\}_D=0;
\end{eqnarray}
\begin{eqnarray}\label{-48S21}
\{\varphi, p\}_D=-(\triangle-V)\delta^3(x-y), \cr
\{\varphi, \varphi\}_D=\{p, p\}_D=0. \qquad
\end{eqnarray}
Then equation of motion for any dynamical variable $z_{dyn}$ can be written as follows:
\begin{eqnarray}\label{-48S25}
\dot z_{dyn}=\{z_{dyn}, H(z_{dyn})\}_D.
\end{eqnarray}
It should be noticed that for the pair $\phi, p$ the Dirac brackets coincide with the Poisson ones. For the
pair $\varphi, p$ the Dirac brackets coincide exactly with the non canonical ones (\ref{-48S12}).

Let us apply the prescription to the model under consideration. The constraints (\ref{-48S19}) imply that either $\phi, p$ or $\varphi, p$ can be chosen to parameterize the dynamical sector of the theory.

Parameterizing it by the pair $\varphi, p$, the equations (\ref{-48S18}) reduce to the Schr\"odinger system
(\ref{-48S2}), (\ref{-48S22}), while the Hamiltonian (\ref{-48S15}) acquires the form (\ref{-48S11}). Notice that $p$ is the conjugate momenta for $\phi$ but not for $\varphi$. Using this Hamiltonian and the Dirac bracket (\ref{-48S21}), Eqs. (\ref{-48S2}), (\ref{-48S22}) can be obtained also according to the rule (\ref{-48S25}). Thus, dynamical variables of this parametrization  of the Lagrangian theory (\ref{-48S13}) represent the real and imaginary parts of the wave function.

Parameterizing dynamical sector by the pair $\phi, p$, the equations (\ref{-48S18}) reduce to the system (\ref{-48S9}), while the Hamiltonian (\ref{-48S15}) acquires the form
\begin{eqnarray}\label{-48S26}
H(\phi, p)=\int d^3x\frac{1}{2\hbar}\left[p^2+[(\triangle-V)\phi]^2\right].
\end{eqnarray}
It is just the Hamiltonian of the theory (\ref{-48S8}).

Hence the classical field theory (\ref{-48S8}) and the Schr\"odinger equation can be identified with two possible parameterizations of dynamical sector of the singular Lagrangian theory (\ref{-48S13}). Specifying parametrization one arrives at either classical or quantum description.

Let us return to the formula (\ref{-48S6}). Our aim now is to show that any solution of the Schr\"odinger equation has the form (\ref{-48S6}). Let $\Psi$$=$$\varphi$$+$$ip$ be solution to the Schr\"odinger equation (\ref{-48S1}). Then there exists the function $\phi(t, x^i)$ such that $\varphi$, $p$, $\phi$, $\pi$ with $\pi=0$ is a solution to the system (\ref{-48S18}), (\ref{-48S19}). Actually, except the first and the last equation, all other equations of the system are already satisfied. The remaining equations with known right hand sides
\begin{eqnarray}\label{-48S27}
\dot\phi=\frac{1}{\hbar}p,
\end{eqnarray}
\begin{eqnarray}\label{-48S28}
(\triangle-V)\phi=-\varphi,
\end{eqnarray}
specify the function $\phi$. Take Eq. (\ref{-48S28}) at $t$$=0$, $(\triangle-V)\phi$$=$$-\varphi(0, x^i)$. The elliptic equation can be solved (at least for the analytic function $\varphi(x^i)$ [16]), let us denote the solution as $C(x^i)$. Then the function
\begin{eqnarray}\label{-48S29}
\phi(t, x^i)=\frac{1}{\hbar}\int_0^t d\tau p(\tau, x^i)+C(x^i),
\end{eqnarray}
obeys the equations (\ref{-48S27}), (\ref{-48S28}). They imply the desired result: the wave function can be presented through the real field $\phi$ and its momenta according to (\ref{-48S6}).
\begin{figure*}[t]
\begin{tabular}{|p{8.5cm}|p{8.5cm}|}
\hline
Electrodynamics & Quantum mechanics\\
\hline
There is the Lagrangian formulation in terms of $A^\mu$ & The same in terms of $\phi$\\
\hline
$A^\mu$ represents the potential for magnetic and electric fields, in the gauge $A^0=0$ one has $\vec B=\nabla\times\vec A$, $\vec E=\partial_t\vec A$& $\Psi=\varphi+ip=-(\triangle-V)\phi+i\hbar\partial_t\phi$\\
\hline
While the Maxwell equations are written in terms of $\vec B$, $\vec E$, the field $\vec E$ is conjugate momenta for $\vec A$ but not for $\vec B$ & While the Schr\"odinger equation is written in terms of $\varphi$, $p$, the field $p$ is conjugate momenta for $\phi$ but not for $\varphi$\\
\hline
Maxwell equations form generalized Hamiltonian system [14, 15] with the Hamiltonian
$\sim\vec E^2+\vec B^2$ & Schr\"odinger equation forms generalized Hamiltonian system with the Hamiltonian $\sim p^2+\varphi^2$\\
\hline
\end{tabular}
\caption{Schr\"odinger field $\phi$ as the wave function potential.}\label{table}
\end{figure*}

\section{Conclusion}
In this work we have associated the classical field theory (\ref{-48S8}) with quantum mechanics of a particle in time independent potential. It has been shown that the Schr\"odinger equation is mathematically equivalent to the second order field equation (\ref{-48S7}) for unique real field $\phi$. Solving the classical theory, one is able to construct the quantum mechanical object, $\Psi$, according to the formula (\ref{-48S6}). The later may be considered as a kind of quantization rule
\begin{eqnarray}\label{-48S30}
\phi_{class}\longrightarrow\Psi_{QM}=D\phi+ip,
\end{eqnarray}
where $p$ is conjugate momenta for $\phi$ and $D$ stands for some differential operator specified by the theory. Origin of the rule has been explained in Section 3: either the classical field  or quantum mechanical equations result from two possible parameterizations of dynamical sector of the singular Lagrangian theory (\ref{-48S13}).

While we have started with quantum mechanics and arrived at a field theory, the procedure can be inverted. It would be interesting to apply it to other (relativistic) field theories.

To conclude with, it should be noticed that the presented formulation implies certain analogy among mathematical structure of the Schr\"odinger equation and the free electrodynamics, see the Fig. 1 above. Roughly speaking, the Schr\"odinger field turns out to be the wave function potential.

\section{Acknowledgments}
Author would like to thank A. Nersessian for encouraging discussions. The work has been supported by the Brazilian foundations CNPq
%(Conselho Nacional de Desenvolvimento
%Científico  e Tecnológico - Brasil)
and FAPEMIG.


\begin{thebibliography}{nn}
\item[$^{(*)}$]E-mail: alexei.deriglazov@ufjf.edu.br ~ On leave of
absence from Dept. Math. Phys., Tomsk Polytechnical University,
Tomsk, Russia.

\bibitem{1} G. Wentzel, Z. Phys. {\bf 38} (1926) 518.

\bibitem{2} H. A. Kramers, Z. Phys. {\bf 39} (1926) 828.

\bibitem{3} L. Brillouin , J. Phys. {\bf 7} (1926) 353.

\bibitem{4} J. Schwinger, Phys. Rev. {\bf 82} (1951) 914.

\bibitem{5} V. P. Maslov and M. V. Fedoriuk, Semi-classical approximation in
quantum mechanics (Dordrecht: D. Reidel Publishing Co., 1981).

\bibitem{6} D. Bohm, Phys. Rev. {\bf 85} (1952) 166; ibid 180.

\bibitem{7} E. Schr\"odinger, Ann. Physik {\bf 81} (1926) 109; see also letters by
Shr\"odinger to Lorentz in: K. Przibram, Briefe z\"ur Wellenmechanik, (Springer-Verlag, Wien, 1963).

\bibitem{8} P.A.M. Dirac, Can. J. Math. {\bf 2}
(1950) 129; Lectures on Quantum Mechanics (Yeshiva Univ., New York, 1964).

\bibitem{9} D. M. Gitman and I. V. Tyutin, Quantization of Fields with
Constraints (Berlin: Springer-Verlag, 1990).

\bibitem{10} A. A. Deriglazov, Phys. Lett. {\bf B 626} (2005) 243-248.
\bibitem{11} A. A. Deriglazov, and K.E. Evdokimov, Int. J. Mod. Phys. {\bf A 15} (2000) 4045.

\bibitem{12} J. M. Souriau, Structure des Systemes Dynamiques (Dund, Paris, 1970).

\bibitem{13} A. A. Deriglazov, Phys. Lett. B530 (2002) 235; JHEP 0303 (2003) 021.

\bibitem{14} W. Pauli (1933), General Principles of Quantum Mechanics (Springer, Heidelberg, 1980).

\bibitem{15} M. Born and L. Infeld. Proc. Roy. Soc. {\bf A 150} (1935) 141.

\bibitem{16} R. Courant and D. Hilbert, Methods of Mathematical Physics (New York:
Interscience Publishers Ltd., 1956).

\end{thebibliography}
\end{document}